\documentclass[twocolumn,10pt]{asme2e}
\usepackage{amsmath,amsfonts}

\usepackage{graphicx}
\graphicspath{{./figs/}}

\newcommand{\divU}{\nabla \cdot u}
\newcommand{\pder}[2][]{\frac{\partial#1}{\partial#2}}
\newcommand{\mbp}{\frac{\mu}{\rho}}
\usepackage{mathtools,xparse}
\DeclarePairedDelimiter{\norm}{\lVert}{\rVert}
\NewDocumentCommand{\normL}{ s O{} m }{%
  \IfBooleanTF{#1}{\norm*{#3}}{\norm[#2]{#3}}_{L_2(D)}%
}

\confshortname{Turbo Expo 2016}
\conffullname{the 2016 ASME Turbomachinery Technical Conference \& Exposition \&\\}

\confdate{13-17}
\confmonth{June}
\confyear{2016}
\confcity{Seoul}
\confcountry{South Korea}

\papernum{GT2016-56689}

\title{DRAFT: Unsteady Adjoint of Pressure Loss for a Fundamental Transonic Turbine Vane}

\author{Chaitanya Talnikar\thanks{Address all correspondence to this author.},
    \hspace{3mm} Qiqi Wang
    \affiliation{
	Aerospace Computational Design Lab\\
	Department of Aerospace and Astrophysics\\
	Massachusetts Institute of Technology\\
	Cambridge, Massachusetts 02139\\
        talnikar@mit.edu, qiqi@mit.edu
    }	
}

\author{Gregory M. Laskowski
    \affiliation{
        Engineering Technologies \\
	GE Aviation \\
	Lynn, Massachusetts, 01910\\
	laskowski@ge.com
    }
}

\begin{document}

\maketitle    

%%%%%%%%%%%%%%%%%%%%%%%%%%%%%%%%%%%%%%%%%%%%%%%%%%%%%%%%%%%%%%%%%%%%%%
\begin{abstract}
    {\it 
        High fidelity simulations, e.g., large eddy simulation are 
        often needed for accurately predicting pressure losses due to wake mixing in turbomachinery applications. 
        An unsteady adjoint of such high fidelity simulations is useful for design optimization in these aerodynamic applications. 
        In this paper we present unsteady adjoint solutions using a large eddy simulation model for a vane from VKI using aerothermal objectives. 
        The unsteady adjoint method is effective in capturing the gradient for a short time interval aerothermal objective, 
        whereas the method provides diverging gradients for long time-averaged thermal objectives. 
        As the boundary layer on the suction side near the trailing edge of the vane is turbulent, 
        it poses a challenge for the adjoint solver. 
        The chaotic dynamics cause the adjoint solution to diverge exponentially from the trailing edge region when solved backwards in time. 
        This results in the corruption of the sensitivities obtained from the adjoint solutions.
        An energy analysis of the unsteady compressible Navier-Stokes adjoint equations indicates
        that adding artificial viscosity to the adjoint equations can potentially dissipate the adjoint energy
        while potentially maintain the accuracy of the adjoint sensitivities. 
        Analyzing the growth term of the adjoint energy provides a metric
        for identifying the regions in the flow where the adjoint term is diverging.
        Results for the vane from simulations performed on the Titan supercomputer are demonstrated.
    }
\end{abstract}

%%%%%%%%%%%%%%%%%%%%%%%%%%%%%%%%%%%%%%%%%%%%%%%%%%%%%%%%%%%%%%%%%%%%%%
\begin{nomenclature}
    \entry{$\bar{p_l}$}{Pressure loss objective}
    \entry{$\bar{p}_{t,l}$}{Mass-averaged stagnation pressure loss on downstream plane}
    \entry{$\bar{p}_{t,in}$}{Stagnation pressure at the inlet}
    \entry{$p_{t,p}$}{Stagnation pressure on downstream plane}
    \entry{$T$}{Time-averaging interval}
    \entry{$S$}{Surface area of boundary}
    \entry{$V$}{Volume of domain}
    \entry{$\rho_p$}{Density on downstream plane}
    \entry{$p_p$}{Pressure on downstream plane}
    \entry{$\gamma$}{Isentropic expansion factor}
    \entry{$M_p$}{Mach number on downstream plane}
    \entry{$\mathbf{x}$}{Position vector}
    \entry{$D$}{Domain for the fluid problem}
    \entry{$\rho$}{Density at a point in the domain}
    \entry{$\mathbf{u}$}{Velocity vector}
    \entry{$c$}{Speed of sound}
    \entry{$E$}{Total energy}
    \entry{$\mu$}{Viscosity coefficient}
    \entry{$\alpha$}{Thermal coefficient}
    \entry{$e$}{Internal energy}
    \entry{$\mathbf{\sigma}$}{Viscous stress tensor}
    \entry{$\mathbf{w}$}{Conservative variables vector}
    \entry{$\mathbf{F}$}{Navier-Stokes convective flux vector}
    \entry{$\mathbf{F^v}$}{Navier-Stokes viscous flux vector}
    \entry{$\bar{J}$}{Time-averaged objective}
    \entry{$J$}{Instantaneous objective}
    \entry{$\mathbf{\hat{w}}$}{Adjoint of conservative variables}
    \entry{$\mathbf{A}$}{Jacobian of convective flux 3-dimensional tensor}
    \entry{$\mathbf{A^v}$}{Jacobian of viscous flux 3-dimensional tensor}
    \entry{$\mathbf{D}$}{Jacobian of viscous flux with respect to
    gradient terms 4-dimensional tensor}
    \entry{$\mathbf{s}$}{Navier-Stokes equations source term vector}
    \entry{$\mathbf{q}$}{Primitive variables vector}
    \entry{$\mathbf{v}$}{Symmetrized variables vector}
    \entry{$\mathbf{S}$}{Transformation matrix from primitive to symmetrized variables}
    \entry{$\mathbf{T}$}{Transformation matrix from conservative to symmetrized variables}
    \entry{$\mathbf{V}$}{Transformation matrix from conservative to primitive variables}
    \entry{$\mathbf{\hat{A}}$}{Primitive flux 3-dimensional tensor for symmetrized variables}
    \entry{$\mathbf{\hat{v}}$}{Adjoint for symmetrized variables}
    \entry{$\mathbf{B}$}{Jacobian of $\mathbf{\hat{A}}$ with respect to symmetrized variables}
    \entry{$\mathbf{\hat{D}}$}{Viscous 4-dimensional tensor}
    \entry{$E_{\hat{v}}$}{$L_2$ norm of adjoint of symmetrized variables}
    \entry{$M, M_1, M_2$}{Growth matrix of adjoint energy}
    \entry{$a,b$}{Scaled speeds of sound}
    \entry{$\sigma_1$}{Maximum singular value of growth matrix}
    \entry{$\lambda$}{Scaling factor for additional adjoint viscosity}
    \entry{$L$}{Length scale of fluid problem}
\end{nomenclature}

%%%%%%%%%%%%%%%%%%%%%%%%%%%%%%%%%%%%%%%%%%%%%%%%%%%%%%%%%%%%%%%%%%%%%%
\section*{INTRODUCTION}
% HIGH FIDELITY SIM
High fidelity simulations like large eddy simulations (LES) 
are essential for accurately simulating turbulent fluid flows.
This is especially true for turbomachinery applications
in which there is a transitioning boundary layer and
flow separation. 
Gourdain \cite{gourdain2012comparison}
compared LES to low fidelity 
methods like Reynolds averaged Navier-Stokes simulations
(RANS) and found that LES predicts heat transfer with a much higher accuracy
when analyzed against experimental data.
Moreover LES is 
becoming more feasible with the growing power of supercomputers.
In just over a decade compute capacity has increased by a factor of 100.
This has enabled high fidelity simulations for fluid problems where the Reynolds number is 
on the order of a million.

% ADJOINT FOR DESIGN
For accomplishing design of turbomachinery components using
LES in a reasonable amount of time it is necessary to
obtain gradients of design objectives to design parameters.
A straightforward method to obtain gradients is to use finite difference,
but the number of simulations that it requires to obtain
the gradient scales linearly with the number of input parameters.
An alternative is to use the adjoint method, which provides
the gradient with respect to a large number of parameters using
just one additional simulation. 
This method has been extensively used for performing
design optimization using steady-state Euler
\cite{jameson1995optimum} or RANS \cite{Lyu2014b} simulations. 
The adjoint method involves
solving a set of equations known as the adjoint equations.
For a time-dependent simulation
like LES an unsteady adjoint method is required in which
the adjoint equations are simulated backwards in time 
to obtain the desired derivatives.
Recently, Economon \cite{economon2013viscous} performed 
unsteady adjoint simulations for a rotating airfoil,
but these were restricted to unsteady laminar fluid flows.

% DIVERGENCE OF ADJOINT
It has been observed in numerous studies 
\cite{wang2013drag,blonigan2012towards} and also 
through simulations conducted for this paper that
for turbulent fluid flows
the unsteady adjoint solution grows exponentially 
when simulated backwards in time. This is due
to the chaotic nature of the turbulent flow field.
From chaos theory \cite{aceves1986chaos}, it is known that the solution
of certain nonlinear partial
differential equations are sensitive to 
perturbations to initial conditions or parameters.
The solutions of the Navier-Stokes equations 
are believed to exhibit
this property and the behaviour has been demonstrated
numerically and experimentally \cite{wilcox1998turbulence}. 
The divergence to infinity of the adjoint field makes it
unusable for computing sensitivities of the objective with
respect to perturbations in input parameters.

% SOLUTION
This paper presents a possible solution to this
problem by controlling the growth of the adjoint field. 
An energy analysis of the unsteady compressible Navier-Stokes adjoint equations 
reflects that a single term contributes to the growth of the energy 
of the adjoint field while another viscous-like term dissipates the adjoint energy. 
This suggests the idea that adding artificial viscosity to 
the adjoint equations 
can dampen the adjoint fields. The maximum singular value of 
the growth term matrix gives an indication of the regions 
in the flow where the adjoint term is diverging at an exponential rate. 
By the addition of minimal artificial viscosity in these regions the growth of 
the adjoint energy can be curbed and at the same time the 
accuracy of the derivatives obtained from the
adjoint solutions can potentially be maintained. 

\section*{PROBLEM SETUP}
\begin{figure}[t]
\begin{center}
    \includegraphics[width=0.25\textwidth]{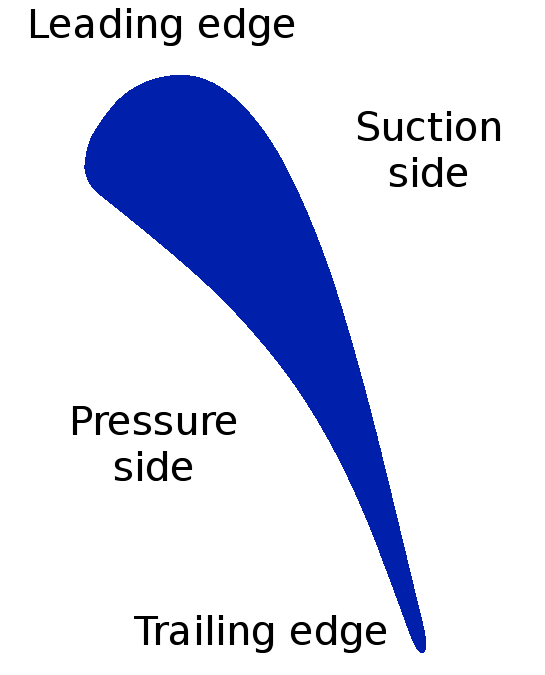}
\end{center}
\caption{TURBINE VANE GEOMETRY}
\label{f:vane_geom} 
\end{figure}

The turbomachinery problem of interest is transonic
flow over a turbine inlet guide vane designed
by researchers at Von Karman Institude (VKI) \cite{arts1992aero} shown
in Figure \ref{f:vane_geom}.

Subsonic flow enters from the inlet on the left side,
accelerates as it goes around the suction side and reaches 
close to the speed of sound near trailing edge of the vane. 
The boundary layer transitions from a laminar to turbulent at the suction side 
near the trailing edge of the vane
as shown in Figure \ref{f:vane_shear_stress}. The point of transition
is highly dependent on the turbulent intensity of the flow at the inlet.
The flow then separates at the trailing edge producing a
turbulent wake. Due to the mixing in the wake there is 
a loss in stagnation pressure of the fluid.

\begin{figure}[t]
\begin{center}
    \includegraphics[width=0.6\textwidth]{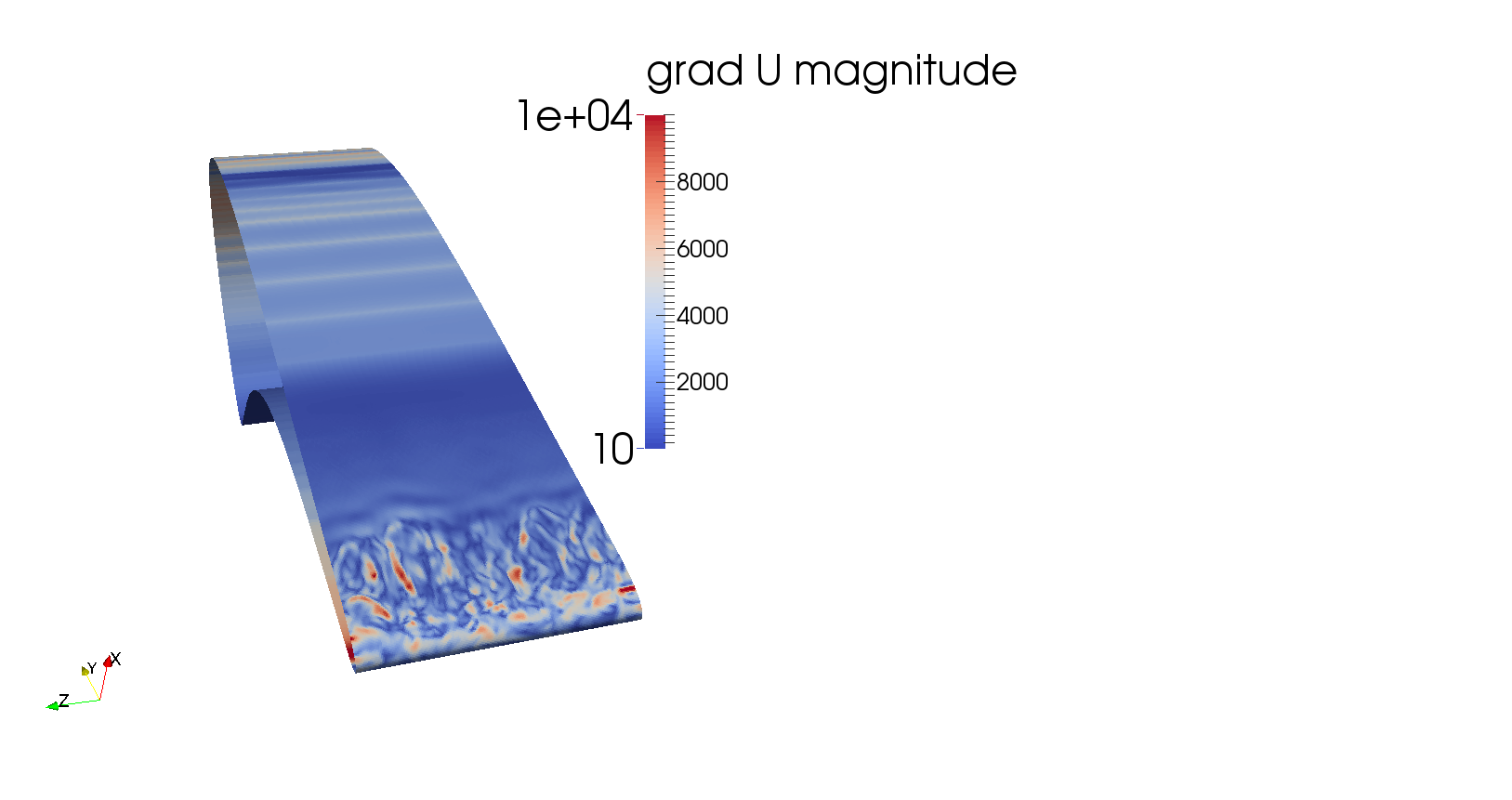}
\end{center}
\caption{CONTOUR PLOT OF SHEAR STRESS ON THE SURFACE OF THE VANE}
\label{f:vane_pt} 
\end{figure}

The vane has a chord length of $67.647$ mm and
the flow from the inlet is at an angle of $\gamma=55$ degrees
to the chord. The vanes are in a linear cascade
and the pitch is $0.85$ times the chord length. 
In the simulation, periodic boundary conditions 
are imposed on the top and bottom. The spanwise extent
of the setup is kept at $10$ mm. 
Numerical studies of this problem have shown that this
is sufficient to capture the dynamics of turbulence \cite{gourdain2012comparison}. 
The vane surface is assumed to be isothermal, realized 
in practice with the help of film cooling holes.

The design objective for this problem is an 
infinite time-averaged and mass-averaged stagnation
pressure loss coefficient ($\bar{p_l}$) $16$ mm downstream of the vane on a surface parallel to the
inlet plane. 
Due to mixing in the wake downstream of the vane there is a large
drop in the stagnation pressure which leads to loss in
performance. Hence, there is an interest in minimizing the
pressure loss.
In practice, the time average for the objective is performed
for an interval which is sufficient to provide a reasonably
accurate estimate of the infinite time average. In this problem
it is equal to the time it takes
for the flow to pass from the inlet to the outlet which comes
to be approximately $2$ ms. The formula for $\bar{p_l}$ is
\begin{align}
    \bar{p_l} &= \frac{\bar{p}_{t,l}}{p_{t,in}} \\
    \bar{p}_{t,l} &= \lim_{T\to\infty}\frac{1}{T}\int_{0}^{T} 
    \frac{\int_S \rho_p(p_{t,in}-p_{t,p})dS}{\int_S \rho_p dS} dt \\
    p_{t,p} &= p_p (1 + \frac{\gamma-1}{2}M_p^2)^{\frac{\gamma}{\gamma-1}}
    \label{e:pressure_loss}
\end{align}

\begin{figure}[t]
\begin{center}
    \includegraphics[width=0.3\textwidth]{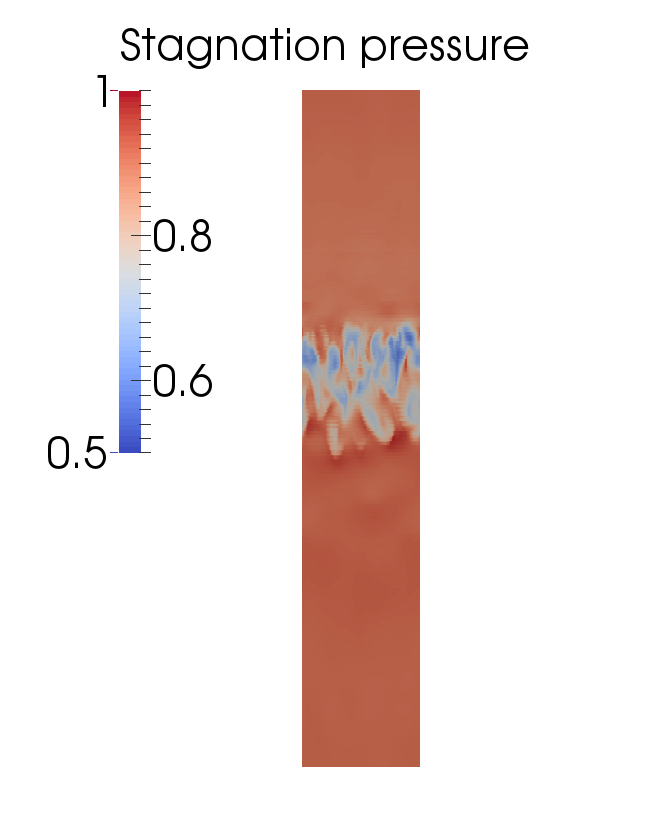}
\end{center}
\caption{STAGNATION PRESSURE ON A VERTICAL
    SPANWISE CROSS SECTION $10$ MM DOWNSTREAM
OF THE TRAILING EDGE OF THE VANE}
\label{f:vane_pt} 
\end{figure}

\subsection*{Time averaged objectives}
\begin{figure}[t]
\begin{center}
    \includegraphics[width=0.5\textwidth]{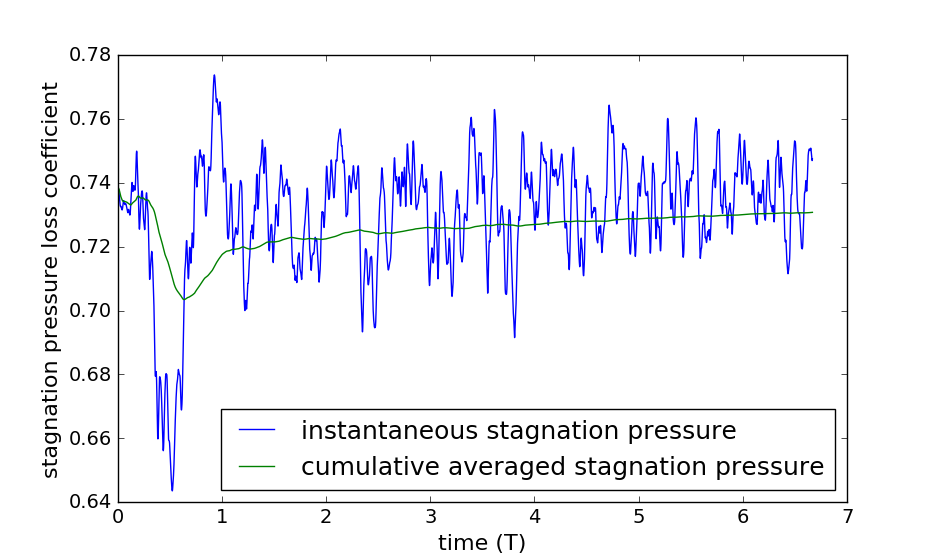}
\end{center}
\caption{INSTANTANEOUS AND TIME-AVERAGED STAGNATION PRESSURE LOSS COEFFICEINT FOR THE VANE.
X-AXIS DENOTES THE TIME NORMALIZED BY THE TIME IT TAKES FOR THE FLOW TO PASS FROM THE
INLET TO OUTLET}
\label{f:vane_pt_history} 
\end{figure}

Observations of turbulent flows indicates that statistics of turbulence 
like time-averaged mean of pressure loss coefficient are well-defined, stable qualtities \cite{lea2000sensitivity,thuburn2005climate}. 
Other than relatively rare cases that exhibits flow hysterisis, these statistics are insensitive to initial conditions. 
In dynamical systems theory, an autonomous system is called ergodic if infinite time averages are independent of initial condition. 
An infinite time average of such ergodic systems is proven to be differentiable to parameters of the system under additional assumptions 
\cite{ruelle2008differentiation,ruelle2009review}. 
This theory is consistent with observations in turbulent flows, in which the statistics are found to depend continuously on parameters when the flow is away from bifurcations
\cite{blonigan2014least}.
Hence, we assume that the pressure loss coefficient is smooth as a function of
inputs like source term perturbations and shape parameters, or
in other words at least the first derivative exists.
Figure \ref{f:vane_pt_history} shows the convergence of the pressure loss coefficient objective as the averaging
interval is increased.

\subsection*{Physics}
The problem can be physically modelled using the 
compressible Navier-Stokes equations with the ideal gas law as 
an approximation for the state equation and appropriate
inlet, outlet and wall
boundary conditions.

\begin{align}
\begin{split}
    \mathrm{In} \hspace{2mm} &\mathbf{x} \in D, t \in [0, T],\\
    &\frac{\partial \rho}{\partial t} + \nabla \cdot (\rho \mathbf{u}) = 0 \\
    &\frac{\partial (\rho \mathbf{u})}{\partial t} + \nabla \cdot (\rho \mathbf{u}\mathbf{u}) + \nabla p = \nabla \cdot \sigma\\
    &\frac{\partial (\rho E)}{\partial t} + \nabla \cdot (\rho E \mathbf{u} + p \mathbf{u}) = \nabla \cdot (\mathbf{u} \cdot \sigma + \alpha \nabla e)  \\
    &\sigma = \mu (\nabla \mathbf{u} + \nabla \mathbf{u}^T) - \frac{2\mu}{3}(\nabla \cdot \mathbf{u})\mathbf{I} \\
    &p = (\gamma - 1)\rho e \\
    &e = E - \frac{\mathbf{u}\cdot \mathbf{u}}{2}
\end{split}
\label{e:navier}
\end{align}

For simulating turbulence, LES provides
a way for resolving the large scale features
of the flow and modelling the small scale structures.
Typically when performing an LES a sub-grid 
scale (SGS) model like Smagorinsky or Vreman \cite{garnier2009large} is used,
but as this problem is simulated using a second
order finite volume method no SGS model is used.
This is because when using a lower order method
the numerical viscosity from the grid might
be of the same order as the subgrid scale viscosity \cite{moeng1989evaluation}.

\subsection*{Numerical methods}

The flow solver is written in Python 
with the help of numpy and scipy libraries
and discretizes the equations over general
unstructured meshes. For time marching a 
strong stability preserving third order Runge-Kutta method \cite{macdonald2003constructing} is used.
An approximate Roe \cite{roe1981approximate} solver is employed for propagating shocks and
discontinuities.
Non reflecting boundary conditions are used for the inlet and outlet. 
Parallelization is accomplished using the Message Passing
Interface (MPI) library.

\begin{figure}[t]
\begin{center}
    \includegraphics[width=0.4\textwidth]{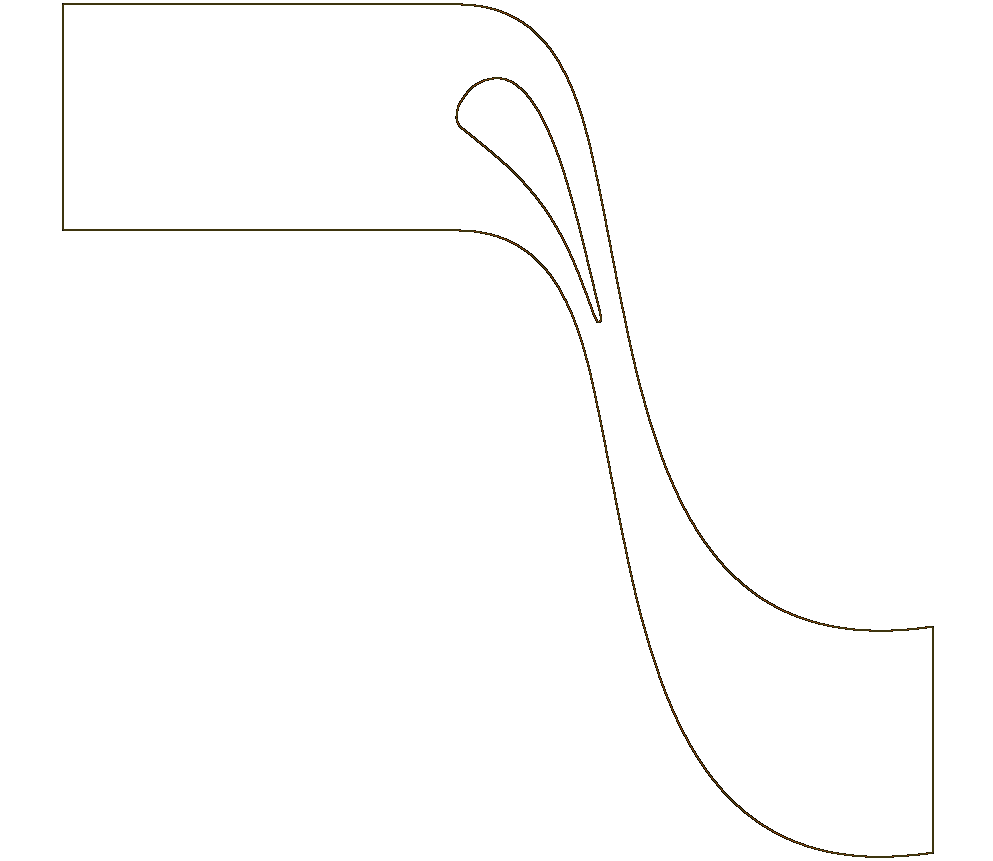}
\end{center}
\caption{TURBINE VANE COMPUTATIONAL DOMAIN}
\label{f:vane_full_geom} 
\end{figure}

The computational domain of the problem is 
shown in Figure \ref{f:vane_full_geom}. 
The simulations are performed on a 2D and 3D
version of the problem. 
The 3D mesh is just
the 2D mesh extended to 3D and
discretized uniformly in the spanwise direction.
The mesh 
is a hybrid structured/unstructured mesh.
It is structured in the inlet, outlet regions
and around the surface of the vane and is
unstructured in the remaining areas.
The smallest cell size is maintained below
$0.5$ mm in regions away from the wall. This
corresponds to a ratio of $6$ between the
smallest eddy and cell size.
To capture all the significant eddies
of the flow (such an LES is called a 
wall resolved LES) near the wall 
the wall normal cell width
has to be below $1$ in terms
of wall units \cite{choi2012grid}. This puts a constraint
on the time step to be of the order of a 
few nanoseconds, tremendously increasing 
the simulation cost. To be able to run
the simulation in a reasonable time frame
the maximum $y+$ is kept at $10$. 
This results in an under resolved LES.
In the future, a wall model might help 
alleviate this problem by allowing the
mesh to have a higher $y+$. For now,
the simulations are run without any 
wall model to get some initial unsteady 
adjoint results.

\section*{UNSTEADY ADJOINT}
The unsteady adjoint method provides a way for computing
derivatives of an objective dependent on the state of a 
system with respect to input parameters
where the state is constrained by a time-dependent partial 
differential equation. 
Rewriting Equation \ref{e:navier} in vector form
\begin{align}
\begin{split}
    \frac{\partial \mathbf{w}}{\partial t} &+ \nabla \cdot \mathbf{F} = \nabla \cdot \mathbf{F^v} \\
    \mathbf{w} &= \begin{pmatrix} \rho \\ \rho \mathbf{u} \\ \rho E \end{pmatrix} \\
    \mathbf{F} &= \begin{pmatrix} \rho\mathbf{u} \\ \rho\mathbf{u}\mathbf{u} \\ (\rho E + p)\mathbf{u} \end{pmatrix} \\
    \mathbf{F^v} &= \begin{pmatrix} 0 \\ \sigma \\ \mathbf{u} \cdot \sigma + \alpha \nabla e \end{pmatrix} \\
\end{split}
\label{e:vector}
\end{align}
Using the Einstein summation notation in the Euclidean space Equation \ref{e:vector}
can be simplified to (with the addition of a source term)
\begin{equation}
    \frac{\partial w_i}{\partial t} + \frac{\partial F_{ij}}{\partial x_j} = \frac{\partial F^v_{ij}}{\partial x_j} + s_i, i=1...5
\end{equation}
Consider a time-averaged objective on the boundary surface ($S$),
$T$ is a large enough time to estimate the infinite 
time average with required accuracy. 
\begin{equation}
    \bar{J} = \int_0^T \int_S J(w_i)dSdt
\end{equation}
The first step in deriving the adjoint equations is to linearize the
governing equation and forming the Lagrangian,
\begin{align}
\begin{split}
    \delta \bar{J} &= \int_0^T \int_S (\frac{\partial J}{\partial w_i}\delta w_i) dSdt \\
                   &+   \int_0^T \int_V\hat{w_i} ( \frac{\partial \delta w_i}{\partial t} + \frac{\partial \delta F_{ij}- \delta F^v_{ij}}{\partial x_j} - \delta s_i)dVdt
\end{split}
\end{align}
Integrating the second term by parts in time and space, 
\begin{equation}
\begin{aligned}
    \delta \bar{J} &= \int_0^T \int_S (\frac{\partial J}{\partial w_i}\delta w_i) dSdt +   \int_V (\hat{w}_{i|T}\delta w_{i|T}-\hat{w}_{i|0}\delta w_{i|0}) dV \\
                   &- \int_0^T \int_V \frac{\partial\hat{ w_i}}{\partial t}\delta w_i dVdt +  \int_0^T\int_S \hat{w_i}(\delta F_{ij} - \delta F^v_{ij})n_j dSdt \\
                   &- \int_0^T \int_V \frac{\partial \hat{w_i}}{\partial x_j} (\delta F_{ij}-\delta F^v_{ij}) dVdt \\
                   &- \int_0^T \int_V \hat{w}_i\delta s_idVdt
\end{aligned}
\end{equation}
Differentiating $F_{ij}, F^v_{ij}$ with respect to $w_k$, $F^v_{ij}$ with
respect to $\frac{\partial w_k}{\partial x_l}$,
\begin{equation}
    A_{ijk} = \frac{\partial F_{ij}}{\partial w_k}, \hspace{10mm}
    A^v_{ijk} = \frac{\partial F^v_{ij}}{\partial w_k}, \hspace{10mm}
    D_{ijkl} = \frac{\partial F^v_{ij}}{\partial \frac{\partial w_k}{\partial x_l}}
\end{equation}
The terms can be rearranged to form,
\begin{equation}
\begin{aligned}
    \delta \bar{J} = &\int_0^T \int_S [(\frac{\partial J}{\partial w_i} + \hat{w_k}(A_{kji}-A^v_{kji})n_j)\delta w_i - \hat{w_i}D_{ijkl}\delta \frac{\partial w_k}{\partial x_l}n_j]dSdt \\ &+   \int_V (\hat{w}_{i|T}\delta w_{i|T}-\hat{w}_{i|0}\delta w_{i|0}) dV \\ &- \int_0^T \int_V (\frac{\partial\hat{ w_i}}{\partial t} + \frac{\partial \hat{w_k}}{\partial x_j} (A_{kji}-A^v_{kji}))\delta w_i dVdt \\
    &+ \int_0^T \int_V \frac{\partial\hat{ w_i}}{\partial x_j}D_{ijkl}\frac{\partial \delta w_k}{\partial x_l} dVdt \\
                   &- \int_0^T \int_V \hat{w}_i\delta s_idVdt
\end{aligned}
\label{e:adjoint_1}
\end{equation}
The last term in Equation \ref{e:adjoint_1} can again be integrated by parts to form
\begin{align}
    \begin{split}
    \int_0^T \int_V \frac{\partial\hat{ w_i}}{\partial x_j}D_{ijkl}\frac{\partial \delta w_k}{\partial x_l} dVdt &= 
    \int_0^T \int_S \frac{\partial\hat{w_i}}{\partial x_j}D_{ijkl}\delta w_k n_l dSdt  \\
    &-  \int_0^T \int_V \frac{\partial}{\partial x_l} (\frac{\partial\hat{ w_i}}{\partial x_j}D_{ijkl})\delta w_k dVdt 
    \end{split}
\end{align}
Using no perturbation in the initial condition, $\delta w_{i|0} = 0$
, the adjoint equation comes out to be
\begin{equation}
     -\frac{\partial\hat{ w_i}}{\partial t} - (A_{kji}-A^v_{kji})\frac{\partial \hat{w_k}}{\partial x_j}  = 
     \frac{\partial}{\partial x_l} (D_{kjil}\frac{\partial\hat{ w_k}}{\partial x_j})
\end{equation}
with boundary condition on the surface
\begin{equation}
(\frac{\partial J}{\partial w_i} + \hat{w_k}(A_{kji}-A^v_{kji})n_j + \frac{\partial\hat{w_k}}{\partial x_j}D_{kjil}n_l)\delta w_i - \hat{w_i}D_{ijkl}\delta \frac{\partial w_k}{\partial x_l}n_j = 0
\end{equation}
And terminal condition $\hat{w}_{i,T} = 0$. Notice that there is a terminal condition which
implies that the adjoint equations have to be integrated backwards in time.
The procedure to compute sensitivity of an objective for a set of perturbations
involves first solving the compressible Navier-Stokes  (primal) equations from
time $t=0$ to $T$. After this, the adjoint equation
is solved backwards in time from $t=T$ to $0$. The adjoint equation requires
the solution of the primal equation at every time $t$. The sensitivity 
due to a perturbation in the source term can be estimated using,
\begin{equation}
    \delta \bar{J} = -\int_0^T \int_V \hat{w}_{i}\delta s_i dVdt
\end{equation}

\subsection*{Implementation}
In practice, the adjoint equation is implemented as a discrete
unsteady adjoint instead of the continuous unsteady adjoint derived in this section.
The discrete adjoint has the advantage that it provides a derivative 
and adjoint sensitivities
that are precise to machine precision when
compared with finite difference sensitivities. 
The discrete adjoint is derived with
the help of automatic differentiation provided
by the Python package Theano \cite{Bastien,Bergstra}.
Additionally,
the checkpointing method is used to provide the adjoint equation with the
necessary primal solutions. This method overcomes the need for storing
all the primal solutions (at every time step) in memory by 
saving periodic snapshots of the 
primal solution on disk. When the adjoint is simulated,
the primal solutions in between two snapshots are obtained
by performing the primal simulation again between the two time points.

\subsection*{Results}
\begin{figure}[t]
\begin{center}
    \includegraphics[width=0.5\textwidth]{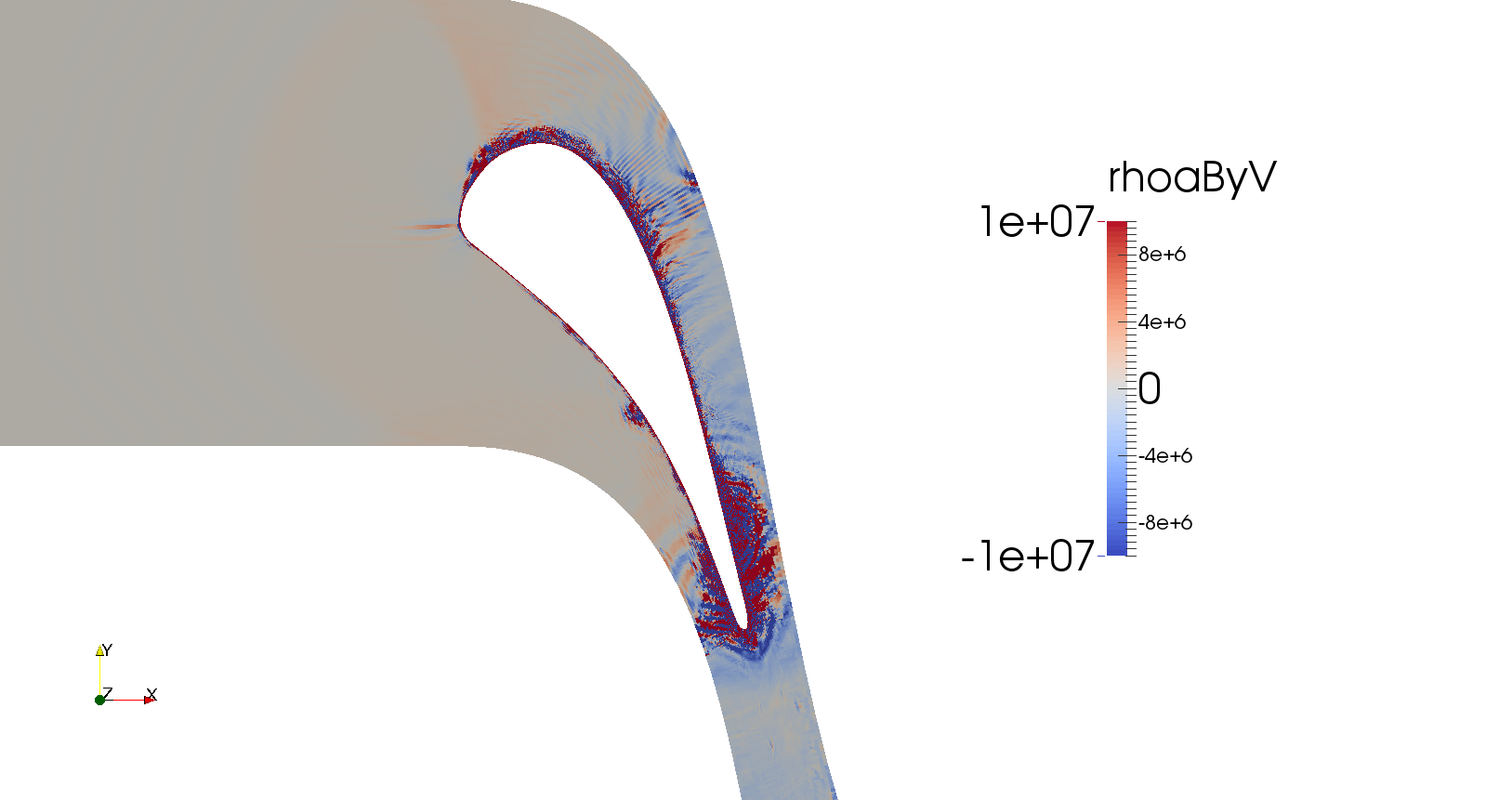}
\end{center}
\caption{A VISUALIZATION OF THE DENSITY ADJOINT FIELD FROM HALFWAY
THROUGH A SHORT TIME 3-D UNSTEADY ADJOINT SIMULATION}
\label{f:vane_2d_rhoa} 
\end{figure}

The unsteady adjoint method is tested on the
turbine vane problem using an aerothermal objective.
A total of $4$ simulations are performed, a 
short time interval and a long time interval simulation 
on each of the 2-D and 3-D turbine
vane problems.  Figure \ref{f:vane_2d_rhoa}
shows a contour plot of the density adjoint field 
for a cross section of the
3-D turbine vane. The adjoint magnitude is large
in the trailing edge region and the leading edge.
The flow is sensitive to perturbations in this area
leading to the high values. The boundary
layer on the suction side of the vane is transitional
and becomes turbulent for high enough inlet turbulence
intensities.

\begin{figure}[t]
\begin{center}
    \includegraphics[width=0.5\textwidth]{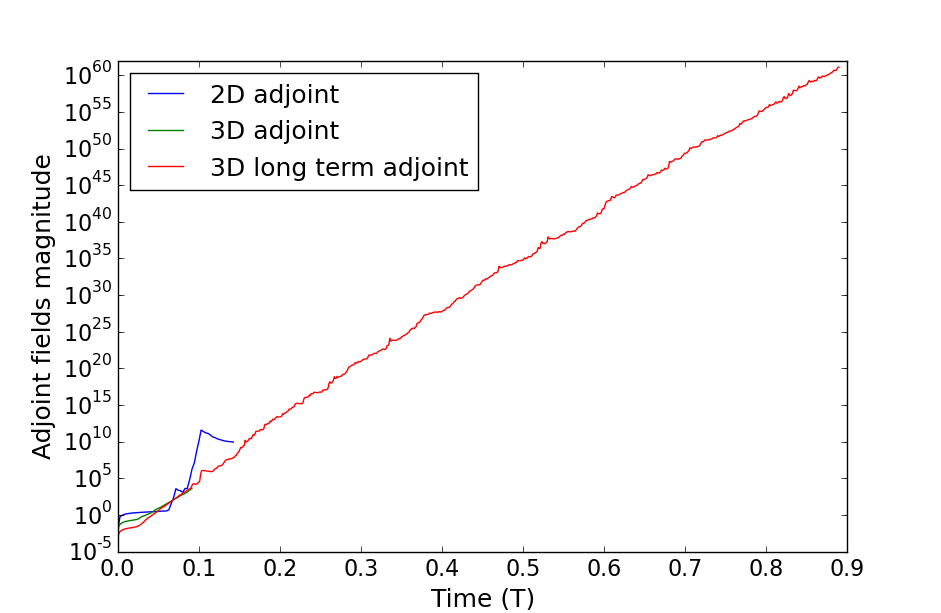}
\end{center}
\caption{GROWTH OF ENERGY NORM OF ADJOINT FIELDS FOR DIFFERENT SIMULATIONS.
$Y$-AXIS SHOWS ENERGY NORM OF A DIMENSIONAL CONSERVATIVE ADJOINT FIELD}
\label{f:vane_energy_growth} 
\end{figure}

The 2-D and 3-D long time interval unsteady adjoints for the turbine vane
diverge exponentially when simulated backwards in
time. 
The actual values of the magnitude
of the adjoint fields in Figure \ref{f:vane_energy_growth} 
are not important,
but what is interesting to note is the
growth rate of the adjoint fields.
The 2-D adjoint diverges at a faster rate
than the 3-D adjoint. This is because the $3^{rd}$ 
dimension provides a way to dissipate the growth in the
adjoint flow field. The long time interval 3-D unsteady adjoint
shows exponential growth throughout the length of the
simulation and it's $L_2$ norm reaches very high magnitudes
of around $10^{60}$.

The high magnitudes of the adjoint fields make
the adjoint sensitivities worthless in the
case of the long time interval 2-D unsteady adjoint ($T=0.1$) and
3-D unsteady adjoint ($T=1$) simulations,
where $T$ is the simulation time normalized
with respect to a single flow through time. 
Table \ref{t:sens} 
gives the actual values of the sensitivities
of the aerothermal objective with respect
to Gaussian shaped source term perturbations in the
conservative flow fields
upstream of the vane. 
The short time interval 2-D and 3-D unsteady adjoint simulations provide
the correct sensitivities. This is because the $L_2$ 
norm of the adjoint fields haven't reached a large
enough magnitude for most of the simulation time.
This result shows that for an objective which only
requires a short time average, the unsteady 
adjoint method can potentially provide the 
correct gradient.

\begin{table}[t]
    \caption{COMPARISON BETWEEN ADJOINT AND FINITE DIFFERENCE SENSITIVITIES}
\begin{center}
    \label{t:sens}
\begin{tabular}{|c|l|l|l|}
\hline
Simulation & T & Finite Diff. & Adjoint \\
\hline
2-D & 0.01 & 4.340e-5 & 4.341e-5 \\
2-D & 0.1 & 0.00695 & 640041 \\
3-D & 0.1 & 0.005904 & 0.005923 \\
3-D & 1 & 0.00710 & $\sim10^{34}$ \\
\hline
\end{tabular}
\end{center}
\end{table}

\section*{STABILIZING ADJOINT EQUATIONS}
Due to the chaotic nature of a turbulent fluid flow, the $L_2$ norm of an unsteady
adjoint solution diverges exponentially when the adjoint equations are solved
backwards in time. This is primarily due to the sensitivity of the solution field 
with respect to small perturbations in the initial flow field or parameters of the
fluid system. 
\begin{equation}
    \normL{\hat{w}} = (\int_V \hat{w}_i\hat{w}_i dV)^{\frac{1}{2}}
\end{equation}

Wang \cite{wang2013drag} performed an energy norm analysis of the adjoint equations
for a fluid flow governed by the incompressible Navier-Stokes equations and 
found out that there two terms which govern the change in adjoint energy of the system.
The first is a growth term which is large in regions where the matrix norm of $\nabla \mathbf{u}$
is large, meaning that regions having a large gradients in velocity contribute to the
divergence of adjoint energy. The second is a dissipation term which tries
to reduce the adjoint energy and is scaled by the viscosity of the fluid. The 
adjoint energy diverges to infinity when the growth term dominates the dissipation
term. This analysis shows that if additional viscosity is added to the 
adjoint equations the dissipation term can limit the growth of the
adjoint field. 

Blonigan \cite{blonigan2012towards} performed numerical experiments of adding uniform 
artificial viscosity to the adjoint equations and was successful
in inhibiting the exponential growth of the adjoint field. But this
also resulted in the corruption of the sensitivities obtained from the
adjoint solution. A potential fix to the latter problem
is to add viscosity 
only in certain regions of the fluid flow where the
adjoint field has a high rate of growth. This idea is 
explored by applying the energy analysis method 
on the unsteady adjoint of the compressible Navier-Stokes equations.

\subsection*{Symmetrization}
Performing the energy analysis on the adjoint of the symmetrized 
Navier-Stokes equations comes out to be more useful than on the
conservative Navier Stokes equations. It can be shown
that if the adjoint energy norm of the symmetrized equations
is bounded then the adjoint energy of the conservative equations
is also bounded. Hence from here forward the focus will be on the
adjoint of the symmetrized equations.

Symmetrization of the Navier-Stokes equations \cite{abarbanel1981optimal} means making the
tensors $A_{ijk}$ and $D_{ijkl}$ symmetric in $i$ and $k$. 
The analysis is performed on the Euler equations, but the
symmetrization procedure works out for the viscous terms too.
Using the quasi-linear form of the Euler equation.
\begin{equation}
    \frac{\partial w_i}{\partial t} + A_{ijk}\frac{\partial w_k}{\partial x_j} = 0
\end{equation}
Symmetrizing by a transformation of the conservative variables to symmetrized
variables, $\delta v_i = T_{ik}\delta w_k$,
\begin{equation}
    T^{-1}_{ik}\frac{\partial v_k}{\partial t} + A_{ijk}T^{-1}_{km}\frac{\partial v_m}{\partial x_j} = 0
\end{equation}
pre-multiplying by $T_{li}$
\begin{equation}
    \frac{\partial v_l}{\partial t} + T_{li}A_{ijk}T^{-1}_{km}\frac{\partial v_m}{\partial x_j} = 0
\end{equation}
$T$ can be chosen such that $\hat{A}_{ljm} = T_{li}A_{ijk}T^{-1}_{km}$ is 
symmetric giving the symmetrized Euler equations,
\begin{equation}
    \frac{\partial v_l}{\partial t} + \hat{A}_{ljm}\frac{\partial v_m}{\partial x_j} = 0
\end{equation}
The symmetrizer from primitive to symmetric variables, $\delta v_i = S_{ik} \delta q_i$ is,
\begin{equation}
    S = \begin{pmatrix}
        \frac{c}{\sqrt{\gamma}\rho} & 0 & 0 & 0 & 0 \\
        0 & 1 & 0 & 0 & 0 \\
        0 & 0 & 1 & 0 & 0 \\
        0 & 0 & 0 & 1 & 0 \\
        -\frac{c}{\rho\sqrt{\gamma(\gamma-1)}} & 0 & 0 & 0 & \sqrt{\frac{\gamma}{\gamma-1}}\frac{1}{\rho c} \\
    \end{pmatrix}
\end{equation}
The transformation from conservative to primitive variables is $\delta q_i = V_{ik} \delta w_i$,
\begin{equation}
    V = \begin{pmatrix}
        1 & 0 & 0 & 0 & 0 \\
        \frac{u_1}{\rho} & \frac{1}{\rho} & 0 & 0 & 0 \\
        \frac{u_2}{\rho} & 0 & \frac{1}{\rho} & 0 & 0 \\
        \frac{u_3}{\rho} & 0 & 0 & \frac{1}{\rho} & 0 \\
        \frac{(\gamma-1)u_{i}u_i}{2} & -(\gamma-1)u_1 & -(\gamma-1)u_2 & -(\gamma-1)u_3 & (\gamma-1) \\
    \end{pmatrix}
\end{equation}
With $T_{ik} = S_{ij}V_{jk}$.

$\hat{A}_{ijk}$ is given by
\begin{equation}
    \hat{A}_{:1:} = \begin{pmatrix}
        u_1 & \frac{c}{\sqrt{\gamma}} & 0 & 0 & 0 \\
        \frac{c}{\sqrt{\gamma}} & u_1 & 0 & 0 & \sqrt{\frac{\gamma-1}{\gamma}}c \\
        0 & 0 & u_1 & 0 & 0 \\
        0 & 0 & 0 & u_1 & 0 \\
        0 & \sqrt{\frac{\gamma-1}{\gamma}}c & 0 & 0 & u_1 \\
    \end{pmatrix}
\end{equation}
\begin{equation}
    \hat{A}_{:2:} = \begin{pmatrix}
        u_2 & 0 & \frac{c}{\sqrt{\gamma}} & 0 & 0 \\
        0 & u_2 & 0 & 0 & 0 \\
        \frac{c}{\sqrt{\gamma}} & 0 & u_2 & 0 & \sqrt{\frac{\gamma-1}{\gamma}}c \\
        0 & 0 & 0 & u_2 & 0 \\
        0 & 0 & \sqrt{\frac{\gamma-1}{\gamma}}c & 0 & u_2 \\
    \end{pmatrix}
\end{equation}
\begin{equation}
    \hat{A}_{:3:} = \begin{pmatrix}
        u_3 & 0 & 0 & \frac{c}{\sqrt{\gamma}} & 0 \\
        0 & u_3 & 0 & 0 & 0 \\
        0 & 0 & u_3 & 0 & 0 \\
        \frac{c}{\sqrt{\gamma}} & 0 & 0 & u_3 & \sqrt{\frac{\gamma-1}{\gamma}}c \\
        0 & 0 & 0 & \sqrt{\frac{\gamma-1}{\gamma}}c & u_3 \\
    \end{pmatrix}
\end{equation}

The adjoint equation for the symmetrized equation comes
out to be slightly different as $\hat{A}_{ljm}$ is not a Jacobian
of the flux term of the symmetric variables.
\begin{align}
\begin{split}
    \delta \bar{J} &= \int_0^T \int_S (\frac{\partial J}{\partial v_i}\delta v_i) dSdt \\
    &+   \int_0^T \int_V\hat{v_i} ( \frac{\partial \delta v_i}{\partial t} + 
    \hat{A}_{ijk}\frac{\partial \delta v_k}{\partial x_j} + \delta \hat{A}_{ijk}\pder[v_k]{x_j})dVdt
\end{split}
\end{align}
Linearizing $\hat{A}_{ijk}$ using $B_{ijkl} = \pder[\hat{A}_{ijk}]{v_l}$ and integrating
by parts in time and space,
\begin{equation}
    \begin{aligned}
        \delta \bar{J} = &\int_0^T \int_S (\frac{\partial J}{\partial v_i}\delta v_i) dSdt \\
    &+ \int_V (\hat{v}_{i|T}\delta v_{i|T}-\hat{v}_{i|0}\delta v_{i|0}) dV - \int_0^T \int_V \frac{\partial\hat{ v_i}}{\partial t}\delta v_i dVdt \\ 
    &+  \int_0^T\int_S \hat{v_i} \hat{A}_{ijk}\delta v_kn_j dS - \int_0^T \int_V \frac{\partial (\hat{v_i}\hat{A}_{ijk})}{\partial x_j}\delta v_k  dVdt
        \\&+ \int_0^T \int_V \hat{v_i}B_{ijkl}\delta v_l\pder[v_k]{x_j}dVdt
    \end{aligned}
\end{equation}
which gives rise to the adjoint equation
\begin{equation}
    -\pder[\hat{v_i}]{t} - \hat{A}_{kji}\pder[\hat{v_k}]{x_j} - (B_{kjil}-B_{kjli})\pder[v_l]{x_j}\hat{v_k} = 0
\end{equation}
The viscous term follows the same derivation
process as done for the conservative adjoint in the previous section with one important difference,
the $\mathbf{F^v}$ term is purely a function of $\nabla \mathbf{v}$ if
$\mbp$ and $\frac{\alpha}{\rho}$ are assumed to be constant
for linearization purposes. This is also known as the frozen viscosity assumption. This implies
that the $A^v_{kji}$ term can be ignored. The $\hat{D}_{ijkl}$ tensor
comes out to be 
\begin{equation}
    \hat{D}_{:1:1} = \begin{pmatrix}
        0 & 0 & 0 & 0 & 0 \\
        0 & \frac{4}{3}\mbp & 0 & 0 & 0 \\
        0 & 0 & \mbp & 0 & 0 \\
        0 & 0 & 0 & \mbp & 0 \\
        0 & 0 & 0 & 0 & \frac{\gamma\alpha}{\rho} \\
    \end{pmatrix}
\end{equation}
\begin{equation}
    \hat{D}_{:2:2} = \begin{pmatrix}
        0 & 0 & 0 & 0 & 0 \\
        0 & \mbp & 0 & 0 & 0 \\
        0 & 0 & \frac{4}{3}\mbp & 0 & 0 \\
        0 & 0 & 0 & \mbp & 0 \\
        0 & 0 & 0 & 0 & \frac{\gamma\alpha}{\rho} \\
    \end{pmatrix}
\end{equation}
\begin{equation}
    \hat{D}_{:3:3} = \begin{pmatrix}
        0 & 0 & 0 & 0 & 0 \\
        0 & \mbp & 0 & 0 & 0 \\
        0 & 0 & \mbp & 0 & 0 \\
        0 & 0 & 0 & \frac{4}{3}\mbp & 0 \\
        0 & 0 & 0 & 0 & \frac{\gamma\alpha}{\rho} \\
    \end{pmatrix}
\end{equation}
\begin{equation}
    \hat{D}_{:1:2} = \hat{D}_{:2:1} = \frac{1}{2}\begin{pmatrix}
        0 & 0 & 0 & 0 & 0 \\
        0 & 0 & \frac{1}{3}\mbp & 0 & 0 \\
        0 & \frac{1}{3}\mbp & 0 & 0 & 0 \\
        0 & 0 & 0 & 0 & 0 \\
        0 & 0 & 0 & 0 & 0 \\
    \end{pmatrix}
\end{equation}
\begin{equation}
    \hat{D}_{:1:3} = \hat{D}_{:3:1} = \frac{1}{2}\begin{pmatrix}
        0 & 0 & 0 & 0 & 0 \\
        0 & 0 & 0 & \frac{1}{3}\mbp & 0 \\
        0 & 0 & 0 & 0 & 0 \\
        0 & \frac{1}{3}\mbp & 0 & 0 & 0 \\
        0 & 0 & 0 & 0 & 0 \\
    \end{pmatrix}
\end{equation}
\begin{equation}
    \hat{D}_{:2:3} = \hat{D}_{:3:2} = \frac{1}{2}\begin{pmatrix}
        0 & 0 & 0 & 0 & 0 \\
        0 & 0 & 0 & 0 & 0 \\
        0 & 0 & 0 & \frac{1}{3}\mbp & 0 \\
        0 & 0 & \frac{1}{3}\mbp & 0 & 0 \\
        0 & 0 & 0 & 0 & 0 \\
    \end{pmatrix}
\end{equation}

If the conservative adjoint field at any point
of time is bounded then the symmetrized adjoint field is bounded
and the vice versa is also true. The sensitivity due to a
perturbation can be computed from either the conservative
adjoint solution or symmetrized adjoint solution.
\begin{align}
    \begin{split}
        \delta \bar{J} &= -\int_0^T \int_V \hat{w}_{i}\delta s_i dVdt \\
                       &=-\int_0^T \int_V \hat{v}_{k}\delta s_k^v dVdt
    =-\int_0^T \int_V \hat{v}_{k}T_{ki}\delta s_i dVdt
    \end{split}
\end{align}
So,
\begin{align}
    \hat{w}_i &= T_{ki} \hat{v}_i \\
    \normL{\mathbf{\hat{w}}} &\le \normL{\mathbf{T}^T} \normL{\mathbf{\hat{v}}}
\end{align}
The transformation matrix consists of bounded component
fields and hence it's matrix $L_2$ norm is bounded. This
implies that if the symmetrized adjoint is bounded then
the conservative adjoint is also bounded.

\subsection*{Energy Analysis}
To study how the adjoint diverges the time 
derivative of the adjoint energy $E_{\hat{v}} = \normL{\hat{v}}$ can be analyzed.
The adjoint energy is basically the sum of the squares of the component-wise
$L_2$ norms. The norms can be summed without dimensional scaling as all the components of the symmetrized
adjoint field have the same dimensions. This is because each of the symmetrized Navier-Stokes
variables has the same dimensions.
\begin{equation}
    -\frac{1}{2}\frac{dE_{\hat{v}}}{dt} = -\frac{1}{2}\frac{\partial}{\partial t}(\int_V \hat{v_i} \hat{v_i}) dV = 
    -\int_V \hat{v_i} \frac{\partial \hat{v_i}}{\partial t} dV
\end{equation}
pre-multiplying the adjoint equation by $\hat{v_i}$ and
integrating over the entire domain,
\begin{align}
\begin{split}
    \frac{dE_{\hat{v}}}{dt} &= \int_V \hat{v_i}(\hat{A}_{kji}\frac{\partial \hat{v_k}}{\partial x_j} +  
    (B_{kjil}-B_{kjli})\pder[v_l]{x_j}\hat{v_k} \\
    &+\frac{\partial}{\partial x_l} (\hat{D}_{kjil}\frac{\partial\hat{ v_k}}{\partial x_j})) dV
\end{split}
\end{align}
using $B_{kjil} = \frac{\partial \hat{A}_{kji}}{\partial v_l}$ the first term can be rewritten as, 
\begin{equation}
\begin{aligned}
    \int_V \hat{v_i}\hat{A}_{kji}\frac{\partial \hat{v_k}}{\partial x_j} dV = 
    &\int_S \hat{v_i}\hat{A}_{kji}\hat{v_k}n_j dS\\
    - &\int_V \frac{\partial \hat{v_i}}{\partial x_j} \hat{A}_{kji}\hat{v_k} dV
    - \int_V \hat{v_i} B_{kjil}\frac{\partial v_l}{\partial x_j}\hat{v_k} dV
\end{aligned}
\end{equation}
The following step requires $A_{kji}$ to be symmetric in $i$ and $k$,
\begin{equation}
\begin{aligned}
    \int_V \hat{v_i}\hat{A}_{kji}\frac{\partial \hat{v_k}}{\partial x_j} dV = 
    \frac{1}{2}(\int_S \hat{v_i}\hat{A}_{kji}\hat{v_k}n_j dS
    - \int_V \hat{v_i} B_{kjil}\frac{\partial v_l}{\partial x_j}\hat{v_k} dV)
\end{aligned}
\end{equation}
The second term in the energy equations is,
\begin{equation}
     \int_V \frac{\partial}{\partial x_l} (\hat{D}_{kjil}\frac{\partial\hat{ v_k}}{\partial x_j}) dV
=
     \int_S \hat{v_i} \hat{D}_{kjil}\frac{\partial\hat{ v_k}}{\partial x_j} n_l dS
     -\int_V \frac{\partial \hat{v_i}}{\partial x_l} \hat{D}_{kjil}\frac{\partial\hat{ v_k}}{\partial x_j} dV
\end{equation}
So, the adjoint energy equation becomes,
\begin{equation}
\begin{aligned}
    \frac{dE_{\hat{v}}}{dt} =
    &\int_V \hat{v_i} [(\frac{B_{kjil}}{2}-B_{kjli})\frac{\partial v_l}{\partial x_j}]\hat{v_k} dV
    -\int_V \frac{\partial \hat{v_i}}{\partial x_l} \hat{D}_{kjil}\frac{\partial\hat{ v_k}}{\partial x_j} dV \\
    + &\int_S \hat{v_i} (\hat{D}_{kjil})\frac{\partial\hat{ v_k}}{\partial x_j} n_l dS
    +\frac{1}{2}\int_S \hat{v_i}\hat{A}_{kji}\hat{v_k}n_j dS
\end{aligned}
\label{e:energy_1}
\end{equation}

Let, 
\begin{equation}
    \begin{aligned}
        M_1 &= \frac{B_{kjil}}{2}\pder[v_l]{x_j} = \pder[\hat{A}_{kji}]{q_l}\pder[q_l]{x_j},  \\
        M_2 &= B_{kjli}\pder[v_l]{x_j} =  \pder[\hat{A}_{kjl}]{q_m}\pder[q_m]{v_i}\pder[v_l]{x_j}
        = \pder[\hat{A}_{kjl}]{q_m}S^{-1}_{mi}\pder[v_l]{x_j}
    \end{aligned}
\end{equation}
The first volumetric term in \ref{e:energy_1} is a quadratic term
in $\mathbf{\hat{v}}$ scaled by the matrix 
 $M = M_1 - M_2$. This is the term that primarily contributes
 to the diverging growth of the adjoint energy.

The second volumetric term in Equation \ref{e:energy_1}
is the dissipation term. Simplification shows that it
is proportional to $\norm{\nabla\hat{v_i}}^2$ scaled
by the viscous coefficient. As the sign in front of
the term is negative, this term reduces the growth of the adjoint
energy.

The boundary terms are quadratic and can
potentially contribute to the growth 
of the adjoint energy. On the inlet and outlet
of the domain, the characteristic boundary conditions are
applied on the basis that the fluid can be assumed to be inviscid
on these boundaries. Hence, the first boundary term
in Equation \ref{e:energy_1} can be ignored. 
Denoting the characteristic Navier-Stokes variables
by $z_i$ and the characteristic adjoint variables by $\hat{z}_i$,
the boundary condition on the inlet and outlet can be written as,
\begin{equation}
    \hat{v}_k \hat{A}_{kji}n_j\delta v_i
    = \hat{z}_k \Lambda_{li}\delta z_i
    = 0
\end{equation}
using the eigendecomposition of $\hat{A}_{kji}n_j = Q_{kl}\Lambda_{lm}Q_{im}$ and
the identities $\delta v_i = Q_{ki} \delta z_i, \hat{z}_i = Q_{ki} \hat{v}_k$.
On the inlet, the characteristic variables coming into
the domain are set, this corresponds to $\delta z_i = 0$
for $i$ where $i^{th}$ characteristic (eigenvalues in eigendecomposition) is negative.
This means that $\hat{z}_j = 0$ for $j$ where the $j^{th}$ characteristic is positive.
Which means that $\mathbf{\hat{v}}$ belongs to the negative eigenspace
of the matrix, implying that the second boundary term
is always negative. Similarly, for the outlet,
the outgoing characteristics are set, which also corresponds 
to the $\delta z_i = 0$ for $i$ where
$i^{th}$ characteristic is negative. Hence,
for characteristic boundaries the second
boundary term is always negative and it
does not contribute to the growth of the adjoint.
Wall boundaries require more analysis as here the
viscous terms are important due to large gradients.
For this paper, the walls are assumed to not 
contribute significantly to the growth of the adjoint term.
This does not mean that the near wall regions are not
important, in fact as we shall see in the next section
they are.

The contribution from the objective source terms to the
adjoint energy is linear and hence they don't directly contribute
to the divergence of the adjoint fields.

Setting $b= \frac{c}{\sqrt{\gamma}}, a = \sqrt{\frac{\gamma-1}{\gamma}}c$.
\begin{equation}
    M_1 = \frac{1}{2}
    %\begin{pmatrix} \divU & \pder[b]{x_1} &  \pder[b]{x_2} & \pder[b]{x_3} & 0\\ 
    %                \pder[b]{x_1} & \divU &  0 & 0 & \pder[a]{x_1} \\ 
    %                \pder[b]{x_2} & 0 &  \divU & 0 & \pder[a]{x_2} \\ 
    %                \pder[b]{x_3} & 0 & 0 & \divU  & \pder[a]{x_3}\\ 
    %                0 & \pder[a]{x_1} & \pder[a]{x_2} & \pder[a]{x_3} & \divU \\ 
    %\end{pmatrix}
    \begin{pmatrix} \divU & \nabla_1 b &  \nabla_2 b & \nabla_3 b & 0\\ 
                    \nabla_1 b & \divU &  0 & 0 & \nabla_1 a \\ 
                    \nabla_2 b & 0 &  \divU & 0 & \nabla_2 a \\ 
                    \nabla_3 b & 0 & 0 & \divU  & \nabla_3 a\\ 
                    0 & \nabla_1 a & \nabla_2 a & \nabla_3 a & \divU \\ 
    \end{pmatrix}
\end{equation}
\begin{equation}
    M_2 = \begin{pmatrix}
        0 & \frac{b}{\rho}\nabla_1 \rho &  \frac{b}{\rho}\nabla_2 \rho & \frac{b}{\rho}\nabla_3 \rho & \frac{\sqrt{\gamma-1}}{2}\divU \\
        0 & \nabla_1 u_1 & \nabla_2 u_1 & \nabla_3 u_1 & \frac{a}{2p}\nabla_1 p \\
        0 & \nabla_1 u_2 & \nabla_2 u_2 & \nabla_3 u_2 & \frac{a}{2p}\nabla_2 p\\
        0 &  \nabla_1 u_3 & \nabla_2 u_3 & \nabla_3 u_3 & \frac{a}{2p}\nabla_3 p \\
        0 & \frac{2}{\gamma-1}\nabla_1 a & \frac{2}{\gamma-1}\nabla_2 a & \frac{2}{\gamma-1}\nabla_3 a & \frac{\gamma-1}{2}\divU \\
    \end{pmatrix}
    %M_2 = \begin{pmatrix}
    %    0 & \frac{b}{\rho}\pder[\rho]{x_1} &  \frac{b}{\rho}\pder[\rho]{x_2} & \frac{b}{\rho}\pder[\rho]{x_3} & \frac{\sqrt{\gamma-1}}{2}\divU \\
    %    0 & \pder[u_1]{x_1} & \pder[u_1]{x_2} & \pder[u_1]{x_3} & \frac{a}{2p}\pder[p]{x_1} \\
    %    0 & \pder[u_2]{x_1} & \pder[u_2]{x_2} & \pder[u_2]{x_3} & \frac{a}{2p}\pder[p]{x_2} \\
    %    0 & \pder[u_3]{x_1} & \pder[u_3]{x_2} & \pder[u_3]{x_3} & \frac{a}{2p}\pder[p]{x_3} \\
    %    0 & \frac{1}{a}\pder[(p/\rho)]{x_1} & \frac{1}{a}\pder[(p/\rho)]{x_2} & \frac{1}{a}\pder[(p/\rho)]{x_3} & \frac{\gamma-1}{2}\divU \\
    %\end{pmatrix}
\end{equation}
Analyzing the matrix $M$, which is the growth matrix, provides a way to find regions in the fluid
flow where the adjoint is diverging. Using Cauchy-Schwartz inequality for matrix/vector norms
\begin{equation}
|\mathbf{\hat{v}}^T \mathbf{M} \mathbf{\hat{v}}| \le \norm{\mathbf{v}}\norm{\mathbf{Mv}} 
\end{equation}
Using the matrix induced $2$-norm and the identity $\norm{\mathbf{Mv}}_2 \le \sigma_{1} \norm{\mathbf{v}}_2$,
$\sigma_1$ being the maximum singular value of $M$
\begin{equation}
    |\mathbf{\hat{v}}^T \mathbf{M} \mathbf{\hat{v}}| \le \sigma_{1}\norm{\mathbf{\hat{v}}}_2^2
\end{equation}
The magnitude of $\sigma_1$ gives an indication of the regions where the adjoint
energy growth is high. It has the
dimensions $\frac{1}{T}$. Additional viscosity is added to the adjoint equations to
curb the divergence of the adjoint field in
the following form,
\begin{equation}
     -\frac{\partial\hat{ w_i}}{\partial t} - (A_{kji}-A^v_{kji})\frac{\partial \hat{w_k}}{\partial x_j}  = 
     \frac{\partial}{\partial x_l} ((D_{kjil} + \lambda \sigma_1 \delta_{ki}\delta_{jl})\frac{\partial\hat{ w_k}}{\partial x_j})
\end{equation}
where $\lambda$ is a scaling factor that is problem specific
and in this work is manually tuned. It has the dimensions
of $L^2$ where $L$ denotes length.

\subsection*{Results}
The stabilized adjoint algorithm is tested on the
2-D turbine vane problem. The objective is the time-averaged
and mass-averaged pressure loss coefficient given by Equation
\ref{e:pressure_loss}. Time-averaging is performed over 
$1/10^{th}$ of a fluid flow through time.

\begin{figure}[t]
    \begin{center}
        \includegraphics[width=0.5\textwidth]{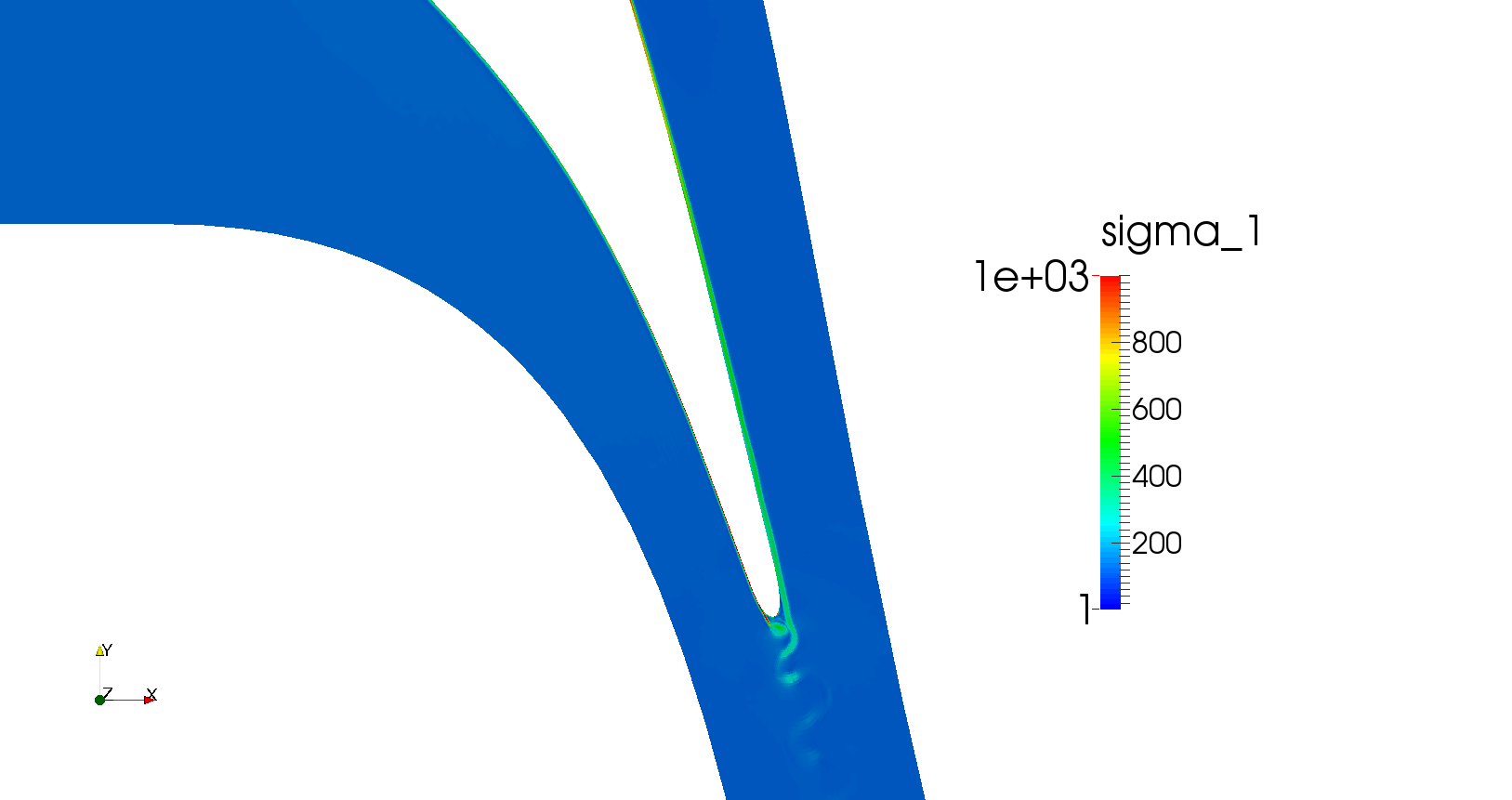}
        \caption{CONTOUR PLOT OF DIVERGENCE INDICATOR $\sigma_1$ FOR THE TURBINE VANE,
        NORMALIZED BY INVERSE OF A SINGLE FLOW THROUGH TIME}
    \label{f:vane_metric}
    \end{center}
\end{figure}

Figure \ref{f:vane_metric} shows the
regions where the maximum singular value $\sigma_1$
of the matrix $M$ is large. As expected,
the region in the boundary layer near the trailing
edge is primarily responsible for the diverging adjoint
and by adding additional dissipation in this region
we can restrict the growth of the adjoint field.

\begin{figure}[t]
    \begin{center}
        \includegraphics[width=0.5\textwidth]{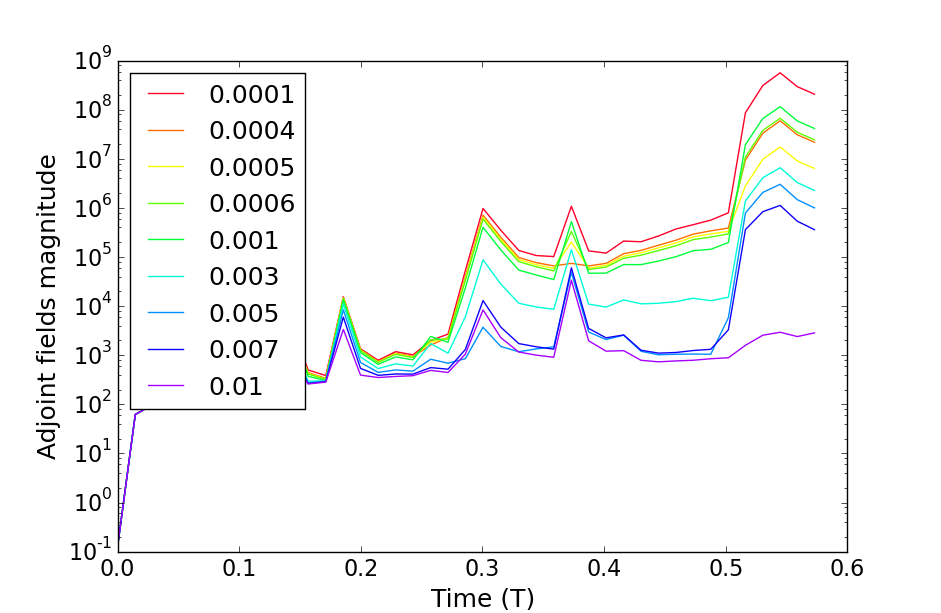}
    \caption{GROWTH OF ENERGY NORM OF ADJOINT FOR VARIOUS SCALING FACTORS.
    X-AXIS IS TIME NORMALIZED BY A SINGLE FLOW THROUGH TIME}
    \label{f:viscous_adjoint_energy}
    \end{center}
\end{figure}

Various values of $\lambda$ are tried from
$\lambda = 10^{-4}$ to $10^{-2}$. 
Figure \ref{f:viscous_adjoint_energy} demonstrates
the growth of energy norm of adjoint fields with time
for the different values of $\lambda$.
When the scaling factor is too low the additional
viscosity does not change the adjoint solution by a significant amount
and the $L_2$ norm of the adjoint fields stays high.
On increasing the scaling factor the magnitude
of the $L_2$ norm of the adjoint fields 
reduces, but still shows exponential growth, 
and this brings the order of the
adjoint sensitivity to match the order of the 
finite difference sensitivity.
Further increase in the scaling factor halts
the exponential growth of the energy norm of the adjoint fields
and reaches an approximate steady state level.
The adjoint sensitivity in this regime agrees
with the finite difference sensitivity within an
error of less than $20\%$. 

%\begin{figure}[t]
%    \begin{center}
%        \includegraphics[width=0.5\textwidth]{vane_error_viscous.png}
%    \caption{ERROR IN ADJOINT SENSITIVITY FOR VARIOUS SCALING FACTORS}
%    \label{f:viscous_adjoint_sens}
%    \end{center}
%\end{figure}
\begin{table}[t]
    \centering
    \begin{tabular}{|c|l|l|}
    \hline
    \multicolumn{3}{|c|}{Sensitivities} \\
    \hline
    Scaling factor ($\lambda$) & Finite Difference & Adjoint \\
    \hline
    0.0001 & 1.71e-4 & 3.44e-3 \\
    0.0004 & 1.71e-4 & 5.42e-4 \\
    %0.0005 & 1.71e-4 & 1.1e-4 \\
    0.0006 & 1.71e-4 & -1.7e-4 \\
    0.001 & 1.71e-4 & -4.5e-4 \\
     0.003 & 1.71e-4 & 1.30e-4 \\
    0.005 & 1.71e-4 & 1.71e-4 \\
     0.007 & 1.71e-4 & 1.59e-4 \\
     0.01 & 1.71e-4 & 1.42e-4 \\
    \hline
    \end{tabular}
    \caption{COMPARISON OF ADJOINT AND FINITE DIFFERENCE SENSITIVITY FOR VARIOUS SCALING FACTORS}
    \label{t:viscous_adjoint_sens}
\end{table}

Table \ref{t:viscous_adjoint_sens} shows
the relative error in adjoint sensitivity with respect
to the finite difference sensitivity for a particular
type of source term perturbation of the compressible
Navier-Stokes equations.
There seems to be an optimal value of $\lambda$
as raising it beyond the point where the
sensitivities match within $0.2\%$ leads
to a slowly increasing error in the sensitivities.
This is due to the fact that too much viscosity 
is being added to the adjoint equations, making
the adjoint sensitivities inaccurate.
This suggests the fact that there is a range of scaling factors where
the adjoint flow field is sufficiently damped
and the sensitivities are reasonably correct.

\section*{CONCLUSION}
The viscous stabilized unsteady adjoint method
provides a promising method for computing adjoint sensitivities
of long time-averaged objectives with respect to
arbitrary perturbations for a turbulent fluid flow.
Next steps for this method is to investigate
how to estimate the scaling factor for a 
specific problem and have different adjoint viscosity $\sigma_1$
for different state variables $\rho, \rho U, \rho E$.
More understanding of the kind of fluid flow
problems where this approach can be beneficial is required.

%%%%%%%%%%%%%%%%%%%%%%%%%%%%%%%%%%%%%%%%%%%%%%%%%%%%%%%%%%%%%%%%%%%%%
%\section*{FOOTNOTES\protect\footnotemark}
%\footnotetext{Examine the input file, asme2e.tex, to see how a footnote is given in a head.}
%
%Footnotes are referenced with superscript numerals and are numbered consecutively from 1 to the end of the paper\footnote{Avoid footnotes if at all possible.}. Footnotes should appear at the bottom of the column in which they are referenced.
%
%
%%%%%%%%%%%%%%%%%%%%%%%%%%%%%%%%%%%%%%%%%%%%%%%%%%%%%%%%%%%%%%%%%%%%%%
% Here's where you specify the bibliography style file.
% The full file name for the bibliography style file 
% used for an ASME paper is asmems4.bst.
\bibliographystyle{asmems4}

%%%%%%%%%%%%%%%%%%%%%%%%%%%%%%%%%%%%%%%%%%%%%%%%%%%%%%%%%%%%%%%%%%%%%%
\begin{acknowledgment}
This research used resources of the Oak Ridge Leadership Computing Facility at Oak Ridge National Laboratory, which is supported by the Office of Science of the Department of Energy under Contract DE-AC05-00OR22725.
\end{acknowledgment}

%%%%%%%%%%%%%%%%%%%%%%%%%%%%%%%%%%%%%%%%%%%%%%%%%%%%%%%%%%%%%%%%%%%%%%
% The bibliography is stored in an external database file
% in the BibTeX format (file_name.bib).  The bibliography is
% created by the following command and it will appear in this
% position in the document. You may, of course, create your
% own bibliography by using thebibliography environment as in
%
% \begin{thebibliography}{12}
% ...
% \bibitem{itemreference} D. E. Knudsen.
% {\em 1966 World Bnus Almanac.}
% {Permafrost Press, Novosibirsk.}
% ...
% \end{thebibliography}

% Here's where you specify the bibliography database file.
% The full file name of the bibliography database for this
% article is asme2e.bib. The name for your database is up
% to you.
\bibliography{asme2e}

\end{document}